\documentclass[conference,10pt]{IEEEtran}
 \IEEEoverridecommandlockouts
\usepackage{cite}
\usepackage{color}
\usepackage[pdftex]{graphicx}
\graphicspath{{fig/}{jpeg/}}
\usepackage[cmex10]{amsmath}
\usepackage{amssymb}
\usepackage{algorithm}
\usepackage{algorithmic}
\usepackage{subcaption}
\usepackage{caption}
\usepackage[outdir=./]{epstopdf}
\usepackage{hyperref}
\usepackage{comment}
\usepackage{bm}
\input{mysymbol.sty}
\usepackage{needspace}

\usepackage{tikz}
\usetikzlibrary{shapes,arrows}
\usepackage{float,environ}
\usetikzlibrary{matrix} 
\usetikzlibrary{decorations.markings} 
\usetikzlibrary{calc} 
\usetikzlibrary{decorations.pathmorphing,snakes, positioning, decorations.pathreplacing}

\usetikzlibrary{fit}

\usepackage{pgfplots}
\usepackage{color}
\usepackage{multirow}

\makeatletter
\newsavebox{\measure@tikzpicture}
\NewEnviron{scaletikzpicturetowidth}[1]{%
  \def\tikz@width{#1}%
  \begin{lrbox}{\measure@tikzpicture}%
  \BODY
  \end{lrbox}%
  \pgfmathparse{#1/\wd\measure@tikzpicture}%
  \BODY
}
\makeatother

\newcommand{\QED}{\hfill\ensuremath{\blacksquare}}

\def\vec2{\text{vec}}

\newtheorem{theorem}{\hspace{0pt}\bf Theorem}

\begin{document} 
%
\title{Optimal WDM Power Allocation via Deep Learning  \\   for Radio on Free Space Optics Systems\thanks{Supported by ARL DCIST CRA W911NF-17-2-0181 and Intel Science and Technology Center for Wireless Autonomous Systems.}
}

\author{\IEEEauthorblockN{Zhan Gao \qquad Mark Eisen \qquad Alejandro Ribeiro}
\IEEEauthorblockA{\textit{Department of Electrical and System Engineering, University of Pennsylvania}, Philadelphia, USA \\
Email: \{gaozhan, maeisen, aribeiro\}@seas.upenn.edu}
}

\maketitle

\begin{abstract}
Radio on Free Space Optics (RoFSO), as a universal platform for heterogeneous wireless services, is able to transmit multiple radio frequency signals at high rates in free space optical networks. This paper investigates the optimal design of power allocation for Wavelength Division Multiplexing (WDM) transmission in RoFSO systems. The proposed problem is a weighted total capacity maximization problem with two constraints of total power limitation and eye safety concern. The model-based Stochastic Dual Gradient algorithm is presented first, which solves the problem exactly by exploiting the null duality gap. The model-free Primal-Dual Deep Learning algorithm is then developed to learn and optimize the power allocation policy with Deep Neural Network (DNN) parametrization, which can be utilized without any knowledge of system models. Numerical simulations are performed to exhibit significant performance of our algorithms compared to the average equal power allocation. 
\end{abstract}

\begin{IEEEkeywords}
Radio on free space optics, deep learning, wavelength division multiplexing, power allocation 
\end{IEEEkeywords}

\section{Introduction}

Modern society has witnessed the appearance of heterogeneous wireless services. These services demand different facilities and operate their own networks respectively, which leads to the high cost and slows down the process of deploying new wireless services. Radio over Fiber (RoF) is put forward as a universal platform that connects multiple radio frequency (RF) signals from different wireless access networks. By placing RF signals on optical carriers, RoF system transmits them through optical fibers without changing their radio formats. It takes advantage of high data-rates, low loss and zero interference, but heavily relies on the deployment of fibers, which may not be available in places like rural areas \cite{Wake2010}.

Free Space Optical (FSO) communication becomes a promising alternative when fibers are not available. With similar advantages as optical fiber communication, it also enjoys license free, easy and inexpensive setup \cite{Chang2006, Andrews2005}. Wireless FSO links get rid of physical restriction of fiber deployment, and is able to transmit RF signals through free space. Therefore, so-called Radio on Free Space Optics (RoFSO) system has been developed recently \cite{Kazaura2010}. However, RoFSO can be seriously affected by FSO channel characteristics, such as weather, turbulence, etc. Different models are proposed for the FSO channel and various techniques are developed to reduce its influence \cite{Andrews2005, Nistazakis2009, Gao2017, Gao2018}.

To improve the performance, Dense Wavelength Division Multiplexing (DWDM) RoFSO system has been developed, as a means of transmitting multiple RF signals simultaneously. It makes it feasible to employ Wavelength Division Multiplexing (WDM) in RoFSO \cite{Bekkali2009, Ciaramella2009}. At the same time, adaptive transmission based on the channel state information (CSI) is proposed to help mitigate channel effects for FSO \cite{Karimi2012, Liu2009} and RoFSO systems \cite{Kim2011, Zhou2015}.

The problem considered in this paper is the optimal power allocation for adaptive WDM transmission in RoFSO systems. According to CSI of all wavelength links, different powers are assigned to different wavelengths to maximize the objective function, subject to power limitation constraints necessary for safe implementation of RoFSO systems. 
The problem is challenging not only because it is both non-convex and constrained, but also because the mathematical system model or estimated CSI may not be accurate in practice. Some model-based algorithms have been developed to handle similar problems \cite{Kim2011,Zhou2015}. These algorithms both employ relaxations to find inexact solutions and are computationally expensive to implement \cite{Zhou2015}. The inherent difficulty makes the application of machine learning appealing, due to both their low complexity and potential for model-free implementation. Deep learning in particular has been applied for resource allocation problems in wireless RF domain in both supervised \cite{sun2017learning} and unsupervised \cite{xu2017deep, eisen2019learning} manners. Such approaches have not yet been explored in FSO or RoFSO systems.

This paper develops two algorithms to solve power allocation for WDM RoFSO. We first formulate the optimal design problem and introduce the RoFSO system model (Section \ref{sec_problem}).  We present the Stochastic Dual Gradient algorithm to solve the problem exactly using the idea of strong duality in \cite{Ribeiro2012} (Section \ref{sec_sdg}). This approach is limited in practice as it is dependent upon system models and requires more computational expense. As a model-free and low complexity alternative, we leverage machine learning techniques in the Primal-Dual Deep Learning algorithm (Section \ref{sec_pddl}). In particular, Deep Neural Networks (DNNs) are used to parameterize the power allocation policy, which are trained with a primal-dual method to solve the resulting constrained learning problem. A model-free implementation is employed using the policy gradient method for cases in which system models are inaccurate or unknown. The strong performance of both algorithms are shown by numerical simulations (Section \ref{sec_numerical_results}).



\section{Problem Formulation}\label{sec_problem}

Radio on Free Space Optics (RoFSO), as a universal platform for heterogeneous wireless services, can transmit RF signals through FSO links in optical networks. The developed Dense Wavelength Division Multiplexing (DWDM) RoFSO system enables the simultaneous transmission of multiple RF signals with WDM technique to increase transmission capacity. Specifically, multimedia RF signals are accessed into RoFSO system and placed on multiple optical wavelength carriers with optoelectronic devices, and then transmitted into free space. At the receiver, optical signals are received through FSO channels, and transferred back to RF signals for users.   

The adaptive transmission is considered when allocating powers to wavelength channels in RoFSO. Based on the channel state information (CSI), different powers are assigned to different wavelengths to maximize the objective function. The exact objective function can be adjusted according to specific situations.

Assume there are $m$ optical wavelengths carrying different transmissions, and each of them are non-overlapping with enough spacing. The CSI is represented by the vector $\bbh = [h_1,...,h_m]$, where each $h_i (i=1,...,m)$ donates the CSI of $i$-th wavelength channel. The allocated power to signal transmitted on the $i$-th wavelength is based upon observed CSI $\bbh$ via a power allocation policy $P_i(\bbh)$. Given the collection of power allocations $\bbP(\bbh) = [P_1(\bbh),...,P_m(\bbh)]$ and current CSI $\bbh$, a channel capacity of $C_i(\bbP(\bbh),\bbh)$ is achieved on the $i$-th wavelength. Note that FSO channel is considered as a fading process with channel coherence time on the order of milliseconds, so we can assume an ergodic and i.i.d block fading process. Since the instantaneous channel capacity tends to vary fast, a long term average $\mathbb{E}_\bbh[C_i(\bbP(\bbh),\bbh)]$ is the more meaningful metric to consider. Additionally, because different wireless services accessed into RoFSO may have different priorities, we consider the weight vector $\bbomega = [\omega_1,...,\omega_m] \ge 0$ to represent such priorities.

There are two natural constraints to be considered in RoFSO power allocation. The first is the expected total power limitation $P_T$ for the FSO base station:
\begin{equation}
\mathbb{E}_\textbf{h} \left[ \sum_{i=1}^m P_i(\bbh) \right] \le P_T.
\end{equation} 
The second is motivated by the eye safety concern in optical transmissions. Specifically, we set a peak power $P_S$ that can be allocated on any single wavelength so that the beam is not dangerous for human eyes in its propagation:
 \begin{equation}
0 \le P_i(\bbh) \le P_S, i=1,...,m.
\end{equation} 

Together, we formulate the optimal power allocation for adaptive WDM transmission in RoFSO systems as the following statistical optimization problem:
\begin{alignat}{3} \label{eq_problem111}
 \mathbb{P}:= &  \max_{\bbP(\bbh)} \ && \sum_{i=1}^m \omega_i \mathbb{E}_\bbh \left[C_i(\bbP(\bbh),\bbh)\right] ,             \\
        &  \st                    \ && \mathbb{E}_\textbf{h} \left[ \sum_{i=1}^m P_i(\bbh) \right] \le P_T,   \nonumber \\
        &                         \ && 0 \le P_i(\bbh) \le P_S, i=1,...,m.   \nonumber%
\end{alignat}

Note that the above problem is formulated without any specific system model. In the proceeding subsection, we discuss channel and capacity models commonly used in the study of RoFSO systems. We then present an exact algorithm to solve \eqref{eq_problem111} in Section \ref{sec_sdg} that relies on such model information, as well as a deep-leraning based alternative algorithm in Section \ref{sec_pddl} that does not.

\subsection{System Model} \label{sec_models}

To mathematically study the FSO channel and RoFSO system, some theoretical models have been put forward in previous researches. For the FSO channel, its effects mainly consist of two parts: the attenuation $h_a$ and the turbulence $h_t$ \cite{Zhou20151}.

The attenuation fading term $h_a$ can be expressed by
\begin{equation}
\begin{split}
&h_a = A(d,\lambda)e^{-\alpha d},\\
&A(d,\lambda)=\frac{A_{TX}A_{RX}}{(d\lambda)^2},
\end{split}
\end{equation}
where $\alpha$ is the attenuation coefficient; $d$ is the transmission distance; $\lambda$ is the wavelength; $A_{TX}$ is the aperture area of transmitter, and $A_{RX}$ is the aperture area of receiver.

As for the turbulence, we use the well-known Log-normal distribution to model the fading term $h_t$, which is considered to be accurate under weak-to-moderate turbulence. Without loss of generality, we can also use other distributions like Gamma-gamma distribution according to different turbulence conditions \cite{Nistazakis2009, Uysal2006}.

The FSO channel can then be modelled as
\begin{equation}
\begin{split}
y = h_a h_t x + n,
\end{split}
\end{equation}
in which $y$ is the received signal; $x$ is the transmitted signal, and $n$ represents the additive Gaussian noise. Therefore, the channel gain (referred as CSI) under this model is expressed by
\begin{equation}
\begin{split}
h = \frac{\vert h_a h_t \vert^2}{N_0}.
\end{split}
\end{equation}

In terms of the RoFSO system with APD photo detector, its performance is commonly evaluated by Carrier to Noise Ratio (CNR), which is modelled as \cite{Dat2009}
\begin{equation}
\begin{split}
CNR = \frac{\frac{1}{2}(OMI\cdot m_prP_r)^2}{RIN\cdot (rP_r)^2+2em_p^{2+F}rP_r+\frac{4KT}{R_f}},
\end{split}
\end{equation}
where $OMI$ donates the optical modulation index; $RIN$ donates the relative intensity noise; $m_p$ is the photodiode gain; $r$ is the photodiode responsivity; $e$ is the electric charge; $F$ is the excess noise factor; $K$ is the Boltzmann's constant; $T$ is the temperature; $R_f$ is the photodiode resistance, and $P_r = Ph$ is the received power at the detector.

With this specific RoFSO system model, the capacity of $i$-th wavelength channel with allocated power $\bbP$ and CSI $\bbh$ can be expressed by
\begin{equation} \label{eq_capacity}
\begin{split}
&C_i(\bbP,\bbh)= \log \left( 1+ CNR_i(\bbP,\bbh) \right) \\
&=\log \left( 1+ CNR_i(P_i,h_i) \right)\\
&= \log \left( 1+ \frac{\frac{1}{2}(OMI\cdot m_prP_ih_i)^2}{RIN \cdot (rP_ih_i)^2+2em_p^{2+F}rP_ih_i+\frac{4KT}{R_f}} \right).
\end{split}
\end{equation}
Note that system parameters are assumed to be same for all wavelength channels.


\section{Stochastic Dual Gradient Algorithm}\label{sec_sdg}

Solving the above optimization problem \eqref{eq_problem111} is challenging due to its non-concave capacity function, functional optimization complexity and the existence of constraints. We first address these challenges by establishing a null duality gap property of \eqref{eq_problem111} and subsequently presenting the Stochastic Dual Gradient (SDG) algorithm to solve. First, for the development of SDG algorithm, we assume that models given in Section \ref{sec_models} are accurate, i.e. the capacity function $C_i(\bbP(\bbh),\bbh)$ in \eqref{eq_problem111} can be computed as in \eqref{eq_capacity}. 

With two constraints in \eqref{eq_problem111}, it is natural to think about working in the dual domain. Let $\mathcal{P} = [0, P_S]^m$ represent the space satisfying the eye safety concern, and introduce the dual variable $\lambda \ge 0$. The Lagrangian of the problem is given by
\begin{equation} \label{eq_lagran}
\begin{split}
\mathcal{L}(\bbP(\bbh),\lambda) & = \sum_{i=1}^m \omega_i \mathbb{E}_\bbh \left[ \log \left( 1+ CNR_i\left( P_i(\bbh), h_i \right) \right) \right] \\
&+ \lambda \left( P_T - \mathbb{E}_\textbf{h} \left[ \sum_{i=1}^m P_i(\bbh) \right] \right). 
\end{split}
\end{equation}
The dual function is then defined as
\begin{equation}
\begin{split}
\mathcal{D}(\lambda) & = \max_{\bbP(\bbh)\in \mathcal{P}} \mathcal{L}(\bbP(\bbh),\lambda). 
\end{split}
\end{equation}
Its corresponding dual problem is to find $\lambda^*$ that minimizes the dual function
\begin{equation} \label{eq_dualprob}
\begin{split}
\mathbb{D} = \min_{\lambda \ge 0} \mathcal{D}(\lambda)=\min_{\lambda \ge 0} \max_{\bbP(\bbh)\in \mathcal{P}} \mathcal{L}(\bbP(\bbh),\lambda). 
\end{split}
\end{equation}

However, the objective function is non-concave and complicated due to the term $CNR_i(P_i(\bbh), h_i)$, which leads it to be a non-convex optimization problem. Solving it in the dual domain then seems to be impossible in principle. Nevertheless note that the key reason here to make the dual method impractical is not the non-convex property but the existence of duality gap indeed, which indicates the loss of optimality if using the dual method. In other words, as long as we can show that this problem does have null duality gap, it is then feasible to be solved in the dual domain. 

Observe that the non-concave objective function is actually inside the expectation expression. We then give the following Theorem 1 according to \cite{Ribeiro2012} to show its null duality gap:
\begin{theorem}
Assume $\mathbb{P}$ and $\mathbb{D}$ donate the optimal solution value of the primal problem \eqref{eq_problem111} and its corresponding dual problem \eqref{eq_dualprob}. If there exists a feasible point $\bbP_0$ satisfying all constraints with strict inequality, and the probability distribution of CSI $\bbh$ contains no point of positive probability, then the duality gap is null:
\begin{equation}
\begin{split}
\mathbb{P}=\mathbb{D}. 
\end{split}
\end{equation}
\end{theorem}

The problem in our case satisfies all conditions of Theorem 1 such that the duality gap is null even if it is a non-convex optimization problem. Then we can directly develop the dual methodology to solve \eqref{eq_problem111} by solving \eqref{eq_dualprob} without any relaxation.
 
The SDG algorithm is put forward based on the above analysis, which iteratively searches for the optimal dual variable $\lambda^*$ from initial $\lambda^0$ and use $\lambda^*$ to compute the optimal power allocation $\bbP^*(\bbh)$. Specifically, at each iteration $k$, SDG consists of two steps:

(1) \emph{Primal variable update.} For the given $\lambda^{k}$ from iteration $k-1$ and CSI $\bbh$, we update the primal variable by maximizing the Lagrangian:
 \begin{equation} \label{eq_priup}
\begin{split}
\bbP^{k+1}(\bbh) &= \argmax_{\bbP(\bbh) \in \mathcal{P}} \mathcal{L}\left(\bbP(\bbh),\lambda^{k}\right)\\
&=\argmax_{\bbP(\bbh) \in \mathcal{P}} \sum_{i=1}^m \omega_i \mathbb{E}_\bbh \left[ \log \left( 1+ CNR_i\left( P_i(\bbh), h_i \right) \right) \right] \\
&\quad \quad \quad+ \lambda^{k} \left( P_T - \mathbb{E}_\textbf{h} \left[ \sum_{i=1}^m P_i(\bbh) \right] \right). 
\end{split}
\end{equation}
Furthermore, both the objective function and constraints separate the use of $P_1(\bbh),...,P_m(\bbh)$ and $h_1,...,h_m$, with no coupling between them. \eqref{eq_priup} can be simplified to
\begin{equation}\label{eq_primalup}
\begin{split}
&P_{i}^{k+1}(\bbh)\\
&= \argmax_{P_i(\bbh) \in [0,P_S]} \omega_i \log \left( 1+ CNR_i\left( P_i(\bbh), h_i \right) \right)- \lambda^{k} P_i(\bbh).
\end{split}
\end{equation}

(2) \emph{Dual variable update.} With $\bbP^{k+1}(\bbh)$ gotten from step (1), we then perform a dual descent method to get $\lambda^{k+1}$:
\begin{equation}
\begin{split}
\lambda^{k+1} &= \left[ \lambda^{k} - \eta^k \nabla_{\lambda} \mathcal{L}(\bm{\theta}^{k+1},\lambda^k) \right]_+ \\
&=\left[ \lambda^{k}-\eta^k \left( P_T - \mathbb{E}_\textbf{h} \left[ \sum_{i=1}^m P_i^{k+1}(\bbh) \right] \right) \right]_+,
\end{split}
\end{equation}
where $\eta^k$ is the stepsize of $\lambda$ at iteration $k$, and $[\cdot]_+$ is due to the non-negativity of $\lambda$. The expectation $\mathbb{E}_{\bbh}[\cdot]$ is computed by the stochastic method with $S$ samples of $\bbh$.

By repeating the above two steps recursively, as $k$ increases, $\lambda^k$ converges to the optimal value $\lambda^*$, and the optimal allocated power of $i$-th wavelength channel $P^*_{i}(\bbh)$ is given by
\begin{equation}
\begin{split}
&P^*_{i}(\bbh)\\
&= \argmax_{P_i(\bbh) \in [0,P_S]} \omega_i \log \left( 1+ CNR_i\left( P_i(\bbh), h_i \right) \right)- \lambda^* P_i(\bbh).
\end{split}
\end{equation}
With the knowledge of accurate system model \eqref{eq_capacity} and SDG algorithm, we can solve the problem \eqref{eq_problem111} perfectly. However, in practice, there exists several problems to discuss:

1. SDG algorithm heavily depends on the exact system model, which means that we need the accurate knowledge of model \eqref{eq_capacity} to perform this algorithm. However, due to the complexity of RoFSO systems, such models may not be accurate in practice.

2. CSI $\bbh$ needs to be estimated at the receiver and feedback to the transmitter. However, the feedback estimated $\widehat{\bbh}$ used in SDG has errors with real $\bbh$ used in the objective capacity function $C_i(\bbP,\bbh)$, which degrades the performance of SDG.

3. In step (1) of SDG, there is not a closed-form solution to the maximization problem to get optimal $P_{i}^{k+1}(\bbh)$, and thus requires time to numerically solve it for each iteration.

These three problems inspires the use of model-free and low-complexity learning algorithms to solve the power allocation problem.


\section{Primal-Dual Deep Learning Algorithm}\label{sec_pddl}

To handle the above limitations of SDG algorithm, we develop the model-free Primal-Dual Deep Learning (PDDL) algorithm, which does not directly use system models but only observed capacity and CSI values. Note that our optimization problem \eqref{eq_problem111} shares the same structure with statistical learning problem. This inspires us to introduce a parametrization $\bm{\theta}\in \mathbb{R}^q$ to represent the power allocation policy $\bbP(\bbh)$ by
\begin{equation} \label{eq_dnn}
\begin{split}
\bbP(\bbh) = \bbPhi(\bbh, \bm{\theta}).
\end{split}
\end{equation}
Substitute \eqref{eq_dnn} into the problem \eqref{eq_problem111}, our purpose then becomes to learn an optimal function $\bbPhi^*(\bbh, \bm{\theta}^*)$ with optimal parametrization $\bm{\theta}^*$, which outputs allocated powers $\bbP^*$ that maximize the objective function.  

In terms of the parametrization, a good choice of $\bbPhi(\bbh, \bm{\theta})$ should provide an accurate approximation for almost any function by changing its parameters $\bm{\theta}$, which can greatly improve the learning performance. Deep Neural Networks (DNNs), widely used in modern machine learning problems, are known to exhibit such strong function approximation ability almost perfectly \cite{hornik1991approximation}. Thus, DNN is a good candidate to be used here. We briefly introduce the architecture of DNN. Assume there are $L$ layers in DNN with $n_1,...n_L$ donating the number of layer units respectively. Each layer is comprised of two parts: linear transform matrix $\bbPi_l \in \mathbb{R}^{n_{l}\times n_{l-1}}$ and non-linear operator $\sigma_l$. The output of $l$-th layer $\bbx_{l} \in \mathbb{R}^{n_{l}}$ can then be obtained by its input $\bbx_{l-1} \in \mathbb{R}^{n_{l-1}}$:
\begin{equation}
\begin{split}
\bbx_{l} = \sigma_l\left( \bbPi_l \bbx_{l-1} \right).
\end{split}
\end{equation}   
Note that the input of DNN $\bbx_0$ is the CSI $\bbh$, and the parametrization $\bm{\theta}$ is the matrices $\{ \bbPi_l \}_{i=1,...,L}$. As for the non-linear operator $\sigma$, various functions can be used, such as ReLu or sigmoid. Besides, note that $\bbtheta$ should belong to the set $\Theta = \{\bbtheta \vert \bbPhi(\bbh, \bbtheta)\in \mathcal{P} \}$ to satisfy the eye safety concern. 

Similar as \eqref{eq_lagran}, the Lagrangian here can be expressed by
\begin{equation}
\begin{split}
\mathcal{L}(\bm{\theta},\lambda) & = \sum_{i=1}^m \omega_i \mathbb{E}_\bbh \left[ C_i(\bbPhi(\bbh, \bm{\theta}),\bbh) \right] \\
&+ \lambda \left( P_T - \mathbb{E}_\textbf{h} \left[ \sum_{i=1}^m \Phi_i(\bbh, \bm{\theta}) \right] \right). 
\end{split}
\end{equation}
And its corresponding dual problem becomes
\begin{equation} \label{eq_dual1}
\begin{split}
\mathbb{D}_{\bm{\theta}} = \min_{\lambda \ge 0} \mathcal{D_{\bm{\theta}}}(\lambda)=\min_{\lambda \ge 0} \max_{\bm{\theta}\in \Theta} \mathcal{L}(\bm{\theta},\lambda). 
\end{split}
\end{equation}
For the above min-max problem with sufficient dense DNN parametrization $\bbtheta$, the duality gap between $\mathbb{P}$ and $\mathbb{D}_{\bbtheta}$ is proportional to the function approximation ability of DNN \cite{eisen2019learning}. Therefore, due to the strong ability of DNN, the duality gap is nearly null. We then develop the PDDL learning algorithm based on \eqref{eq_dual1}, which updates primal variable $\bm{\theta}$ and dual variable $\lambda$ simultaneously at every iteration using first order gradients. The ultimate purpose is to search for a local stationary point $(\bbtheta^*, \lambda^*)$ which satisfies KKT conditions. Specifically, at each iteration $k$, we follow two steps:

(1) \emph{Primal variable update.} For a given $\lambda^{k}$ from iteration $k-1$ and CSI $\bbh$, we update the primal variable $\bm{\theta}$ by
\begin{equation} \label{eq_priup1}
\begin{split}
\bm{\theta}^{k+1} &= \bm{\theta}^{k} + \delta^k \nabla_{\bm{\theta}} \mathcal{L}(\bm{\theta}^{k},\lambda^{k})\\
& = \bm{\theta}^{k} + \delta^k \nabla_{\bm{\theta}} \mathbb{E}_\bbh \left[ \sum_{i=1}^m \omega_i C_i( \bbPhi(\bbh, \bm{\theta}^{k}),\bbh) \right. \\
& \left. + \lambda^{k} \left( P_T - \sum_{i=1}^m \Phi_i(\bbh, \bm{\theta}^{k}) \right) \right],
\end{split}
\end{equation}
where $\delta_k$ is the stepsize of $\bm{\theta}$ at iteration $k$, and the last equation is because of the linearity of the expectation.

(2) \emph{Dual variable update.} Once we get $\bm{\theta}^{k+1}$, the dual variable $\lambda$ is updated by a similar way
\begin{equation} \label{eq_dualup1}
\begin{split}
\lambda^{k+1} &=\left[ \lambda^{k} - \eta^k  \left( P_T -\mathbb{E}_\bbh \left[ \sum_{i=1}^m \Phi_i(\bbh, \bm{\theta}^{k+1}) \right] \right) \right]_+.
\end{split}
\end{equation}

Observed from \eqref{eq_priup1}, the update of primal variable requires not only computing the gradient of capacity function $C_i( \bbPhi(\bbh, \bm{\theta}^k),\bbh)$, but also taking expectation $\mathbb{E}_\bbh[\cdot]$ of this gradient w.r.t. the distribution of $\bbh$. Either of them may be hard to know in practice, which makes the above algorithm useless. However, so-called policy gradient method used in reinforcement learning provides a good solution for these problems. It can be used to calculate the gradient for functions with the form of $\mathbb{E}_\bbh [f(\bbPhi(\bbh,\bm{\theta}),\bbh)]$, where $f$ is an unknown function. Actually, it calculates a stochastic and model-free approximation for $\nabla_{\bm{\theta}} \mathbb{E}_\bbh [f(\bbPhi(\bbh,\bm{\theta}),\bbh)]$ \cite{Sutton2000}. 

In policy gradient method, the power allocation policy $\bbPhi(\bbh, \bm{\theta})$ is considered to be drawn from a distribution with a delta density function $\pi_{\bbh,\bm{\theta}}(\bbP)=\delta(\bbP - \bbPhi(\bbh, \bm{\theta}))$, and then we can rewrite
\begin{equation} \label{eq_grad}
\begin{split}
\nabla_{\bm{\theta}} \mathbb{E}_\bbh [f(\bbPhi(\bbh,\bm{\theta}),\bbh)] = \mathbb{E}_{\bbh,\bbP}[f(\bbP,\bbh) \nabla_{\bm{\theta}} \log \pi_{\bbh,\bm{\theta}}(\bbP)],
\end{split}
\end{equation}
in which $\bbP$ is a random realization drawn from the distribution $\pi_{\bbh,\bm{\theta}}(\bbP)$. However, calculating $\nabla_{\bm{\theta}} \log \pi_{\bbh,\bm{\theta}}(\bbP)$ of a delta density function still requires the knowledge of $f$. To handle this problem, the delta function can be approximated by Gaussian distribution centered around $\bbPhi(\bbh,\bm{\theta})$. And its mean and variance are given by the output features of DNN. Then we can estimate $\nabla_{\bm{\theta}} \mathbb{E}_\bbh [f(\bbPhi(\bbh,\bm{\theta}),\bbh)]$ by using \eqref{eq_grad} without knowing $f$. In addition, we take $S$ samples and average them when computing $\mathbb{E}_{\bbh,\bbP}[\cdot]$ to reduce the stochastic error:
\begin{equation}
\begin{split} \label{eq_policy}
\widetilde{\nabla_{\bm{\theta}}} \mathbb{E}_\bbh [f(\bbPhi(\bbh,\bm{\theta}),\bbh)] = \frac{1}{S}\sum_{j = 1}^S f(\bbP_j,\bbh_j) \nabla_{\bm{\theta}} \log \pi_{\bbh_j,\bm{\theta}}(\bbP_j),
\end{split}
\end{equation}
where $\bbh_j$ is a sampled CSI and $\bbP_j=[P_{j,1},...,P_{j,m}]$ is a corresponding realization drawn from the distribution $\pi_{\bbh_j,\bm{\theta}}(\bbP)$. So with \eqref{eq_policy}, we can compute the gradient in step (1) by
\begin{equation} \label{eq_polgrad}
\begin{split}
&\widetilde{\nabla_{\bm{\theta}}} \mathcal{L}(\bm{\theta},\lambda)\\
&=\widetilde{\nabla_{\bm{\theta}}} \mathbb{E}_\bbh \left[ \sum_{i=1}^m \omega_i C_i( \bbPhi(\bbh, \bm{\theta}),\bbh) \right. \left. + \lambda \left( P_T - \sum_{i=1}^m \Phi_i(\bbh, \bm{\theta}) \right) \right] \\
&= \frac{1}{S}\sum_{j = 1}^S \left\{ \left[ \sum_{i=1}^m \omega_i C_i( \bbP_j,\bbh_j) \right. \right.\\
& \left. \left. + \lambda \left( P_T - \sum_{i=1}^m P_{j,i} \right) \right] \nabla_{\bm{\theta}} \log \pi_{\bbh_j,\bm{\theta}}(\bbP_j) \right\}.
\end{split}
\end{equation}

Therefore, the primal variable update of PDDL algorithm can be completed by using \eqref{eq_polgrad} without any knowledge of system model $C_i( \bbP,\bbh)$ or CSI distribution but only their observations, which makes PDDL model-free. By replacing $\nabla_{\bm{\theta}} \mathcal{L}(\bm{\theta}^k,\lambda^k)$ with $\widetilde{ \nabla_{\bm{\theta}}} \mathcal{L}(\bm{\theta}^k,\lambda^k)$ in \eqref{eq_priup1}, PDDL is summarized in the following Algorithm 1.

{\linespread{0.9}
\begin{algorithm}[b] \begin{algorithmic}[1]
\STATE \textbf{Input:} Initial primal and dual variables $\bbtheta^0, \bblambda^0$
\FOR [main loop]{$k = 0,1,2,\hdots$}
      \STATE Draw CSI samples $\{ \bbh \}$ of batch size $S$, and get their corresponding $\{ \bbP \}$ according to $\text{DNN}_{\bbtheta^k}$, $\pi_{\bbh_j,\bm{\theta}^k}(\bbP)$
      \STATE Obtain observations of capacity values $C_i( \bbP,\bbh)$ at current samples of step 3
      \STATE Compute the policy gradient $\widetilde{ \nabla_{\bbtheta}} \mathcal{L}(\bbtheta^k, \lambda^k)$ by \eqref{eq_polgrad}
      \STATE Update the primal variable by \eqref{eq_priup1}  \\ 
	$\bm{\theta}^{k+1} = \bm{\theta}^{k} + \delta^k \widetilde{ \nabla_{\bm{\theta}}} \mathcal{L}(\bm{\theta}^k,\lambda^k) \nonumber $
	\STATE Update the dual variable by \eqref{eq_dualup1}\\
	$\lambda^{k+1} = \left[ \lambda^{k} - \eta^k \nabla_{\lambda} \mathcal{L}(\bm{\theta}^{k+1},\lambda^k) \right]_+ \nonumber $
\ENDFOR

\end{algorithmic}
\caption{Primal-Dual Deep Learning Algorithm}\label{alg:learning} \end{algorithm}}

 
 \section{Simulation Results}\label{sec_numerical_results}

In this section, we perform numerical simulations to exhibit the performance of SDG and PDDL algorithms, and show their validity by comparing with the average equal power allocation policy. Wavelengths in 1520nm-1595nm are used in simulations with 5nm guard band between adjacent wavelengths. 

Note that although our PDDL learning algorithm is model-free, we are doing numerical simulations not physical experiments. CSI samples $\{\bbh \}$ and their corresponding channel capacities $\{ C_i( \bbPhi(\bbh, \bm{\theta}),\bbh) \}$ cannot be observed here. We then still use the system model to compute them, but in reality we can directly get them from the real system in experiments without the need of any theoretical model. In addition, due to the separable use of $P_i$ and $h_i$ in both objective function and constraints, we construct $m$ independent DNNs for $m$ wavelength channels, and each DNN has three hidden layers with 20, 10 and 5 units respectively. ReLU function is utilized as the non-linear operator $\sigma$. Furthermore, the truncated Gaussian distribution is used as power policy distribution $\pi_{\bbh,\bbtheta}(\bbP)$ in policy gradient method, which constrains generated powers $\bbP$ inside $\mathcal{P}$ to satisfy the eye safety concern. The outputs of DNNs are used as means and standard deviations of $\pi_{\bbh,\bbtheta}(\bbP)$.  

\begin{figure}[t]
\centering
\includegraphics[width=0.47\linewidth, height=1\textheight, keepaspectratio]{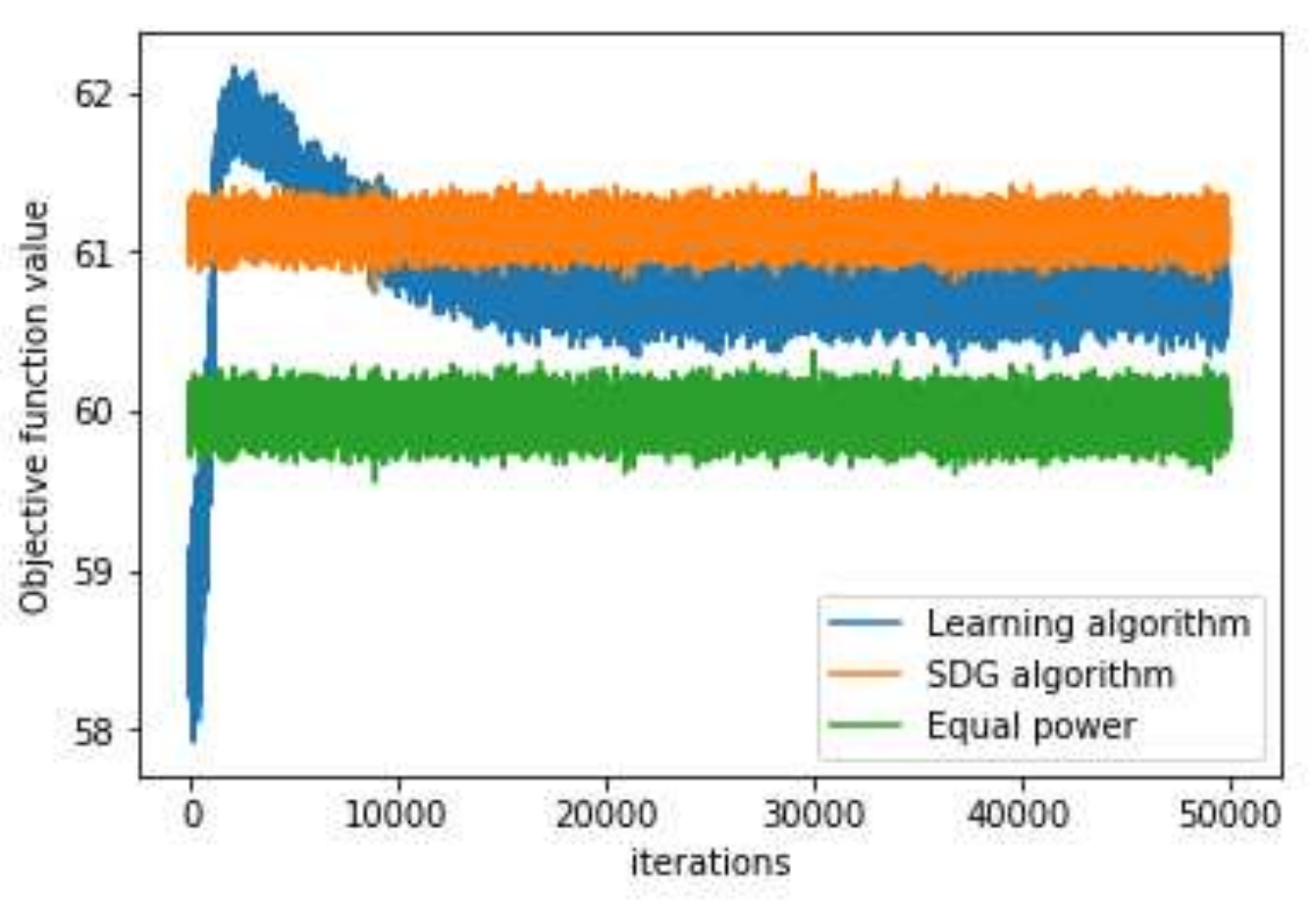} 
\quad
\includegraphics[width=0.47\linewidth, height=1\textheight, keepaspectratio]{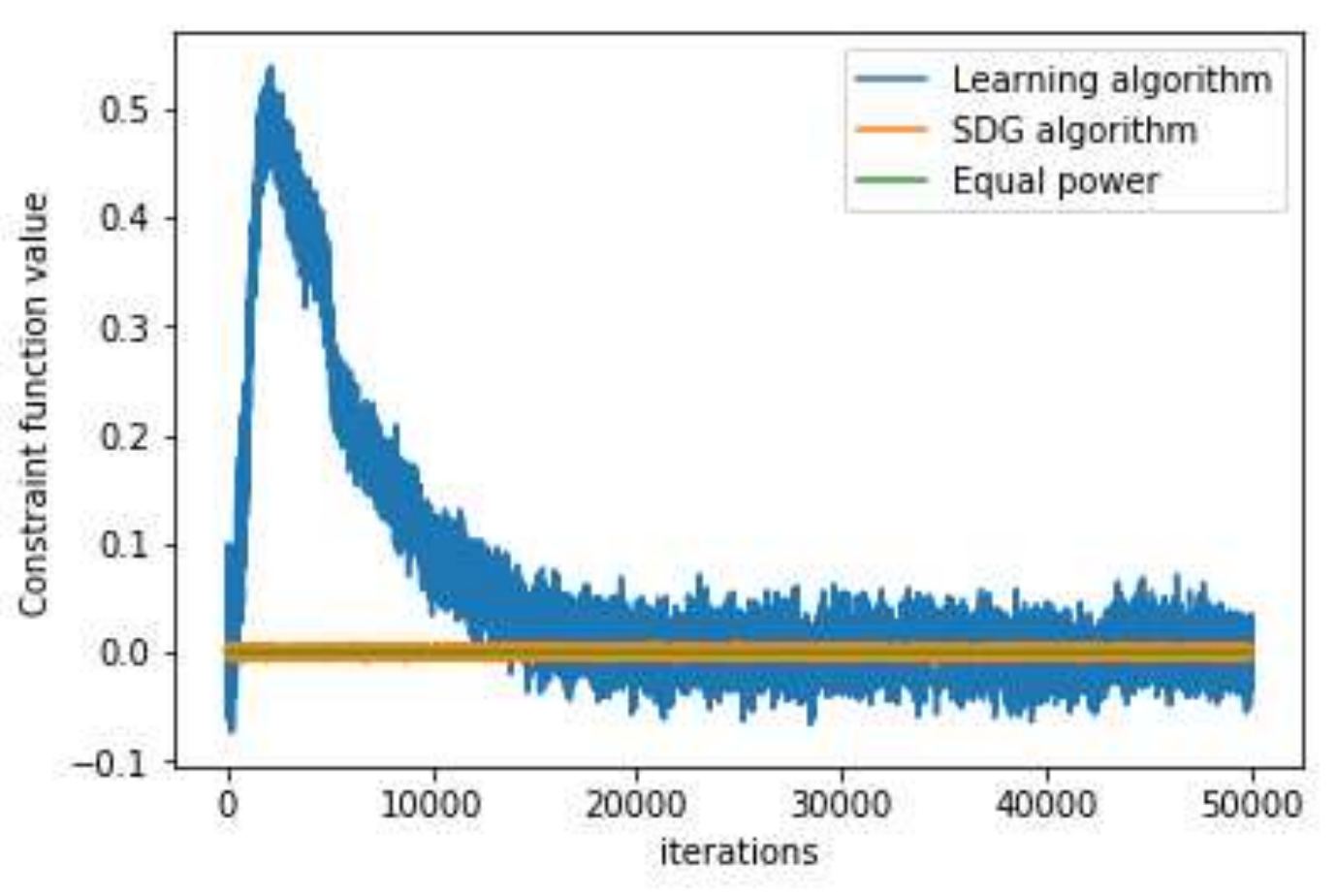} 
\quad

\caption{The objective function value (left) and the constraint function value (right) over learning iterations of three policies for $8$ wavelength multiplexing.}\label{fig_simple_results}
\end{figure}

Fig. 1 shows the performance of three policies for $8$ wavelength multiplexing. Weights $\bbomega$ are drawn randomly from $0$ to $1$, and other default parameters are set as: $P_T = 1.2W$; $P_S = 0.3W$; $m_p=5$; $OMI=15\%$; $r=0.8$; $RIN=-140dB/Hz$; $T=300K$; transmitter aperture diameter $D_{tx}=0.05m$; receiver aperture diameter $D_{rx} = 0.1m$; $d=1000m$. Note that these parameter values are taken as an example to show our algorithms' performance, which can be adjusted based on specific systems and experiments. It can be seen from the left figure that the objective function values achieved by SDG and PDDL learning algorithms converge as iteration increases, and the performance of them outperforms the equal power policy. Similarly, the right figure plots the constraint function values with the increasing of iteration. The values eventually converge to $0$ for both of our algorithms, which indicates the feasibility of their optimal solutions. Besides, note that the model-based SDG that solves the problem exactly exhibits the best performance, which matches with our analysis. On the other hand, the objective value achieved by model-free PDDL converges closely to that of SDG, which validates the near perfect performance of PDDL. Moreover, PDDL can be used without any knowledge of system models, which is particularly useful when FSO system models are unknown, inaccurate, or too complicated to deal with, while SDG cannot handle such situations. Additionally, SDG requires to numerically solve a local maximization problem \eqref{eq_primalup} for every $\lambda^k$ and $\bbh$. Though it is not too hard since it is one-dimensional and with no constraint, SDG is still computationally more expensive than PDDL.

 \begin{figure}[t]
\centering
\includegraphics[width=0.47\linewidth, height=1\textheight, keepaspectratio]{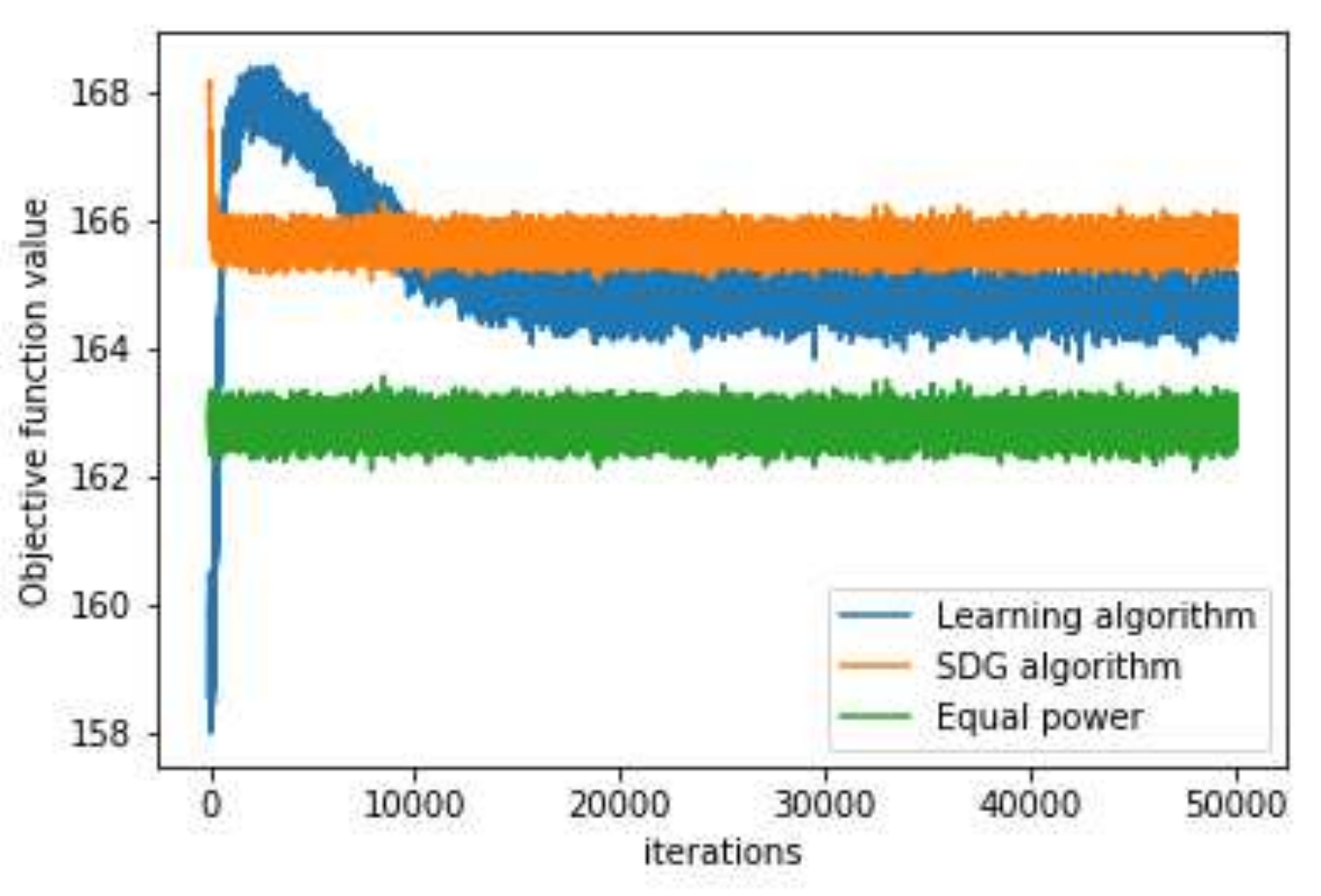} 
\quad
\includegraphics[width=0.47\linewidth, height=1\textheight, keepaspectratio]{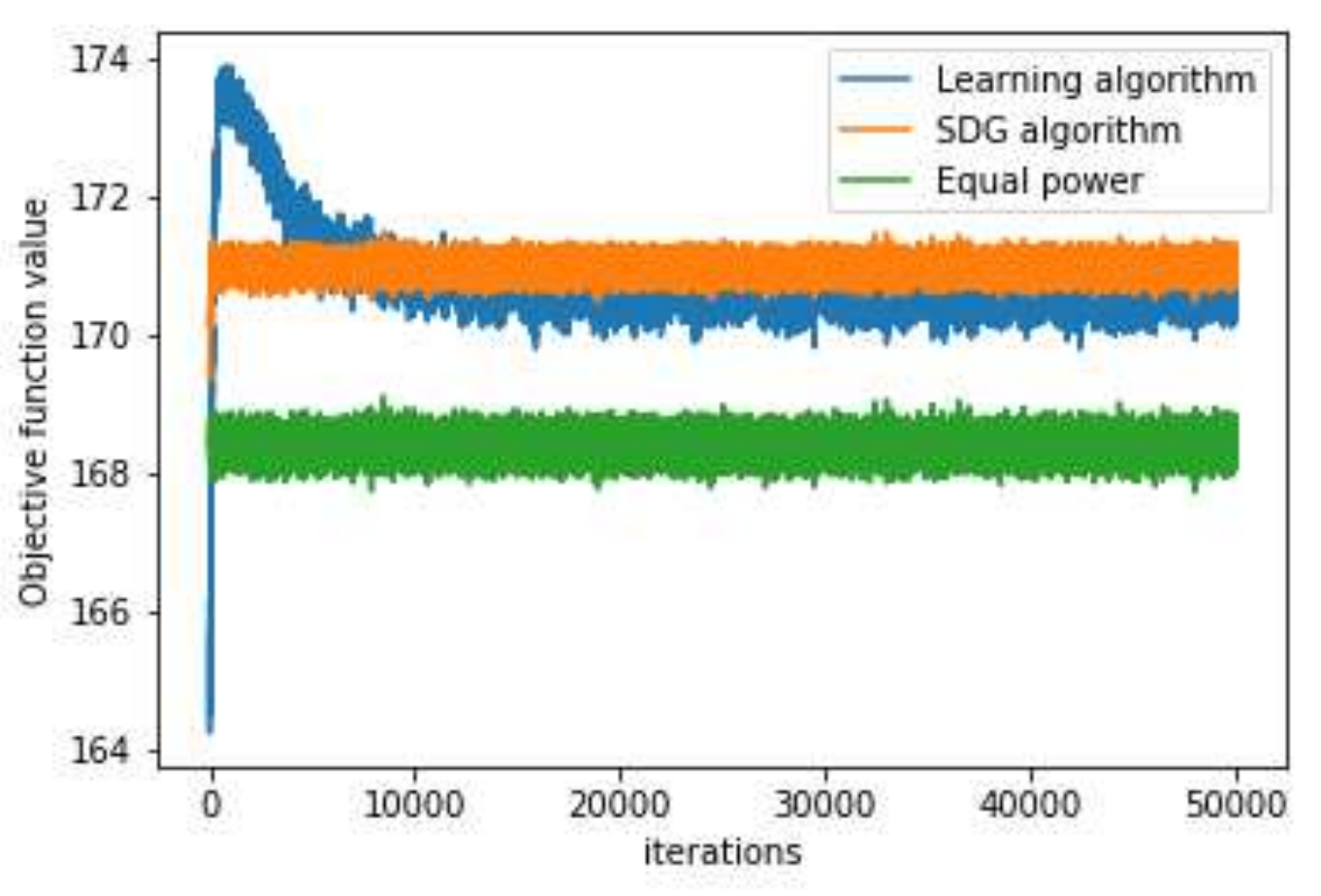} 
\quad

\caption{The objective function value over learning iterations of three policies for $16$ wavelength multiplexing with $P_T = 2.4W, P_S = 0.3W$ (left) and $P_T = 4.0W, P_S = 0.5W$ (right).}\label{fig_simple_results}
\end{figure}

In the left figure of Fig. 2, we depict the performance of three policies for $16$ wavelength multiplexing with $P_T = 2.4W, P_S = 0.3W$. Results show both SDG and PDDL learning algorithms perform well for larger WDM systems, and the advantage of PDDL compared to the equal power policy becomes bigger. The right figure plots their performance for $16$ wavelength multiplexing with larger power settings $P_T = 4W, P_S = 0.5W$, which means there is more space for algorithms to manipulate powers. We can see that PDDL performs better in this case and converges roughly the same value as the exact solution found by SDG.

 
 \section{Conclusion} \label{sec_con}
This paper investigates the challenging problem of optimal power allocation for WDM transmission in RoFSO systems. Two algorithms are developed to adaptively assign powers to different wavelength channels based on CSI. By showing the null duality gap, we first present the model-based Stochastic Dual Gradient algorithm, which is able to solve the problem exactly but heavily relies on the system model and CSI estimation accuracy. The model-free Primal-Dual Deep Learning algorithm is then developed to overcome the shortcomings of SDG. Specifically, it parameterizes the power allocation policy with Deep Neural Networks and learns optimal parameter values by updating primal and dual variables simultaneously. Policy gradient method is applied to compute updating gradients without using the knowledge of system or channel models. Numerical simulations are performed to show that both of our algorithms outperform the equal power policy. The model-free PDDL learning algorithm presented in this paper has wide applications for problems in FSO networks and communications, where FSO systems are sophisticated to model and turbulent channels are complicated to estimate.

\bibliographystyle{IEEEbib}
\bibliography{FSO_learning,wireless_learning}

\end{document}